# WASHtsApp – A RAG-powered WhatsApp Chatbot for supporting rural African clean water access, sanitation and hygiene

*Working Paper*

Simon KLOKER[1,2] [0000-0003-3653-9836], Alex Cedric LUYIMA[1], Matthew BAZANYA[1]
[1]*Ndejje University, Uganda*
*Email: {skloker, aluyima, mbazanya}@ndejjeuniversity.ac.ug*
[2]*Coworkers, Germany*

**Abstract:** This paper introduces WASHtsApp, a WhatsApp-based chatbot designed to educate rural African communities on clean water access, sanitation, and hygiene (WASH) principles. WASHtsApp leverages a Retrieval-Augmented Generation (RAG) approach to address the limitations of previous approaches with limited reach or missing contextualisation. The paper details the development process, employing Design Science Research Methodology. The evaluation consisted of two phases: content validation by four WASH experts and community validation by potential users. Content validation confirmed WASHtsApp's ability to provide accurate and relevant WASH-related information. Community validation indicated high user acceptance and perceived usefulness of the chatbot. The paper concludes by discussing the potential for further development, including incorporating local languages and user data analysis for targeted interventions. It also proposes future research cycles focused on wider deployment and leveraging user data for educational purposes.

**Keywords:** WASH, chatbot, RAG, LLM, WhatsApp, rural development, SDG6

## 1. Introduction

According to the United Nations' sixth Sustainable Development Goal (SDG 6), everyone is entitled to access clean, safe, affordable water and basic sanitation, including latrines and waste collection [1]. Better sanitation and water management can boost a country's economic growth and reduce poverty. This is particularly important in low-to-lower-middle-income countries like Uganda, where unsafe water sources and poor sanitation contribute to an estimated one million deaths globally every year [2].

Uganda's current National Population and Housing Census 2024 displayed part of this problem. 7% of the households in Uganda do not have toilets, and in some areas, up to 60%. 50% have only "unimproved sanitation facilities". 19% of the households do not have access to clean water, and almost 48% still collect their water from public taps or boreholes. While this is already an improvement from the past, exposure to water-borne diseases is still high and causes significant risks to individual health and overall economic growth.

To address this crisis, health education, which involves a combination of learning experiences based on sound theories, can empower individuals, groups, and communities to make informed health decisions [3]. Such education can come via digital media. A study in Tanzania found that even media access itself is positively correlated to WASH practices [4]. In recent years, chatbots have become an increasingly used "digital media" tool for health-related educational tasks. During COVID-19, over 61 chatbots emerged in 30 countries to provide advice, conduct risk assessments, and educate while maintaining social distancing [5]. However, most of these chatbots, including those in healthcare, relied on simple decision-tree designs. Additionally, their overall evaluation was often lacking [5], [6]. A prime example of existing solutions is the WashKaro chatbot, developed in India during the COVID-19 pandemic and was particularly effective in combating misinformation and promoting WASH awareness. At the time, misinformation about the pandemic spread rapidly. To address this, WashKaro was designed to deliver accurate information aligned with WHO recommendations in understandable local languages [7]. Another example is Sirona, a Menstrual Health Management Chatbot available on WhatsApp [8].

Developments in the last few years have significantly raised the quality and applicability of chatbots. Progress, especially concerning the technology of Large Language Models (LLMs), has empowered conversations with AI chatbots on almost human levels [9]. Several publications have already discussed the use of such LLMs as ChatGPT to tackle a multitude of the Sustainable Development Goals [10], also SDG6: "Clean water and sanitation", mainly regarding its possible deployment in the area of education [10], [11], [12]. However, to date, no academic discussions or implementations of such chatbots have been reported. Additionally, real-world implementations must address key LLM limitations, such as "hallucination", before becoming truly useful. Hallucination occurs when LLMs, operating in data-sparse contexts or unfamiliar contexts, fabricate information and present it with high confidence (confidence conundrum) [13]. This is particularly dangerous in the context of health-related issues [14]. Fine Tuning and Retrieval Augmented Generation (RAG) approaches are possible measures to address these issues, whereby RAG seems to outperform Fine Tuning in low knowledge contexts [13], [15]. RAG is an approach to provide the relevant knowledge to context-specific questions to the LLM alongside the prompt, including the question. Therefore, the content is split into chunks, transforming each into a vector using sentence embedding. The user question is then used to search this vector store and retrieve relevant chunks that are provided to the LLM. RAG was also successfully implemented to contextualise LLM output in a Ugandan setting [16].

The current study aims to develop a prototype of a solution to serve rural Ugandan areas with advice regarding WASH Principles in the form of a chatbot. There is mutual agreement that this is a pressing issue in Uganda. Although our chatbot is the of its type in Uganda, we additionally employ and test two new approaches to distinguish us from previous implementations: (1) Our chatbot does not serve as an individual app but via WhatsApp, reachable on a Ugandan Mobile Phone number to increase ease-of-use, acceptance and reach. (2) Our chatbot answers questions using a RAG-powered approach to increase the quality of answers and conversational flexibility. Such a prototype can be used to evaluate and display the value of such a Service to the community and other relevant stakeholders. This is especially important, as many factors of the user group are yet unknown, like IT affinity, their exact knowledge requirements and user behaviour overall. It can also assess to what extent RAG technology is suitable to contextualise the LLM answers to the rural Ugandan setting. To the best of our knowledge, this is the first artifact of this type in this context scientifically captured and evaluated.

The remainder of the paper is structured as follows: The Methodology section describes the two validation approaches to ensure content validity and community acceptance. Thereafter, the Technology Description section presents the developed artifact. The subsequent Development section describes the artifact's architecture and technological approaches employed. The Results section presents the evaluation results for both. The Business Benefits section and the Conclusion highlight and place the results, lessons learned, and the paper's contribution in their appropriate context.

## 2. Methodology

WASHtsApp was developed following the design science research methodology, according to Vaishnavi & Kuechler (2004) [17]. This paper details our initial Design Science Circle focused on creating a prototype artifact. To ensure relevance, we assess whether the proposed technology design can (1) provide high-quality answers to WASH-related questions and (2) be perceived as applicable by the target user group. To achieve this, we employ a rigorous two-step validation process: First, the validation of the content, and second, the validation of the acceptance of technology.

We sourced questions about WASH principles from the internet to validate the chatbot's content. We focused on job interview questions for prospective WASH Officers, assuming that questions people would ask the app would be similar to those they would ask persons knowledgeable in clean water, sanitation and hygiene matters in rural African areas. We also incorporated a questionnaire assessing general WASH knowledge in rural Indian settings. Overall, we merged five sources (four regarding WASH officer interview questions and the other mentioned) and removed duplicates or quasi-duplicates and questions that are not related to WASH specifically[1]. Also, we excluded clearly irrelevant questions but erred on the side of inclusion when uncertain. The final question set comprised 93 questions, adapted to an open-ended format to mimic real-world usage, avoiding multiple-choice questions. In the next step, we asked the chatbot these questions and recorded their answers. Average answer time was at 5.04 seconds (min: 3 seconds, max: 13 seconds). These question-and-answer pairs were presented to four experts for assessment on the scale provided in Table 1. Experts were sourced from Ndejje University, Uganda (2), University of Applied Sciences Rottenburg, Germany (1) and The Ayin Project, United States (1). None of the experts were part of the research team. Experts assessed the question-answer pairs under researcher supervision, ensuring adequate time for thorough evaluation. Participation was voluntary, and no compensation was offered. Although one expert would be enough to assess the chatbot's answers, we opted for four experts, even from different backgrounds, to adequately cater for ambiguous or context-sensitive questions.

*Table 1 Scale to assess the WASHtsApp provided answers by the experts.*

| Scale Item | Explanation (provided to the Expert) |
|---|---|
| Perfect | The correct answer with the proper explanation depth to provide enough context. |
| Sufficient | A correct answer that at least answers the question. |
| Not Sufficient | An answer that is not wrong but does not answer the question sufficiently. (Mark this also when the chatbot states that it cannot answer the question.) |
| Wrong | A wrong answer. |
| I don't know | Mark this if you don't know whether the answer is correct or not. |

The complete questionnaire, including the briefing, can be found in the online appendix.

To validate the acceptance of the technology, we conducted a convenience sample of individuals from our target context. We introduced WASHtsApp and allowed participants to interact with it on our smartphones or their own devices, as preferred. After a few minutes of interaction, we asked them to provide feedback, both verbally and with a short questionnaire. The questionnaire included limited demographic information and six items adapted from the Technology Acceptance Model (TAM) [18]. We adapted the item wording from Venkatesh & Davis (2000) [19], replacing "the system" with "WASHtsApp" and adjusting the context to "my everyday life". The items are listed in Table 2. We applied a 5-point Likert scale ranging from "Fully agree" to "Fully disagree". The questionnaire was available in print or online, depending on the participants' preferences. The questionnaire is provided in the online appendix.

---
[1] The interview questions included questions regarding mere knowledge of the statistical methodology and general, abstract information on problems regarding water access worldwide.

*Table 2 Items of the community validation questionnaire. The items were displayed in a mixed order.*

| Construct | Item |
|---|---|
| Intention to Use | Assuming I have access to WASHtsApp, I intend to use it. |
| | Given that I have access to WASHtsApp, I predict that I would use it. |
| Perceived Usefulness | Using WASHtsApp enhances my ability to stay healthy. |
| | I find WASHtsApp to be useful in my everyday life. |
| Perceived Ease of Use | My interaction with WASHtsApp is clear and understandable |
| | I find WASHtsApp to be easy to use. |

The suggested following Design Cycles are briefly outlined in Section 7.

## 3. Technology Description

WASHtsApp is a RAG-powered chatbot that responds to WhatsApp messages. It is currently accessible via WhatsApp at +256 760 299775. Figure 1 illustrates an interaction with WASHtsApp. The technology presents itself to the user in the same way as any other WhatsApp conversation. The technical implementation of WASHtsApp is outlined in section 4. The AI is trained to answer questions about WASH principles and related topics. For other inquiries, it politely declines to answer. A small website linked to the phone number outlines the Privacy and Data Protection policy.

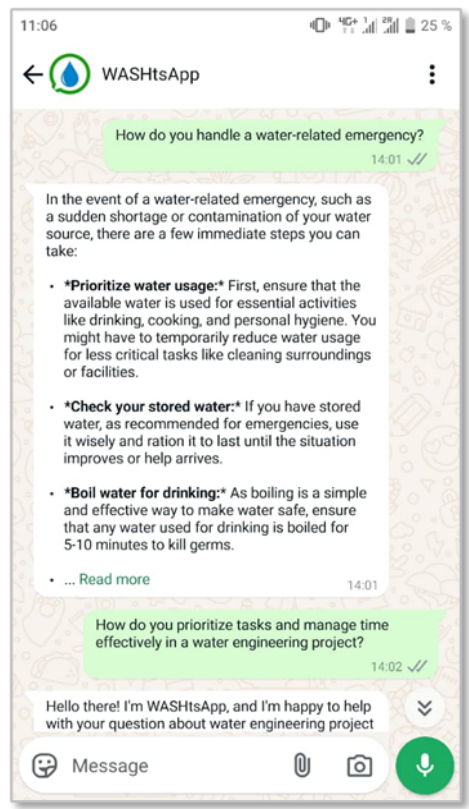

*Figure 1 Screenshot of a communication with WASHtsApp in WhatsApp.*

## 4. Developments

WASHtsApp comprises five main components: User Interface, Backend Connector, Backend, RAG Component, and Database.

The **User Interface** leverages a WhatsApp Business Account registered with a Ugandan mobile number. This comes with several advantages:
- Many users are familiar with the interface and usage, and there is no learning curve.
- Many users have installed WhatsApp, and there is no inertia in installing a new app.
- The user interface of WhatsApp is optimised for user experience.

The **Backend Connector** utilises the WhatsApp Business API Webhook. There are several ways to connect WhatsApp to a custom Backend server. Different approaches vary in cost and feature set. While the WhatsApp Business API lacks built-in chatbot features and a user-friendly communication interface, it offers two key advantages to other approaches:
- We do not need to have a smartphone running to keep the communication up and are, therefore, at least on this side independent of power and network (compared to, e.g., AutoResponder App)
- The cost is lower as if we would include another third-party platform that would also ask for its share (e.g., compared to Twilio Platform)

A simple Flask application (Python) deployed on Google Cloud App Engine serves as the **Backend**. The Backend is rather slim and performs only some very basic checks. It primarily prepares incoming messages for the RAG Component (parsing webhook calls and constructing retrieval chains) and formats its output for the WhatsApp Business API. We selected the Google Cloud App Engine for hosting due to its rapid deployment, cost-effectiveness (pay-per-use model), and robust performance. Furthermore, Google Cloud App Engine eliminates the need for server maintenance, simplifying overall application management.

The **RAG Component** leverages an LMM and Sentence Embedding Model from Cohere, accessed through LangChain and the Cohere API. A document[2] [20] about WASH Principles in rural settings was transformed to a vector store using the Cohere Sentence Embedding Model, which persisted on the Backend. The backend then builds a retrieval chain with the vector store and the cohere LLM to retrieve a response to the request.

Google Firebase Store was used as a database for reasons similar to those of the Google Cloud App Engine.

## 5. Results

*5.1 Content Validation*

The results of the expert assessments of WASHtsApp's answers are summed up in Figure 2. We did not average the expert assessments, as the scale does not allow it, but only added the total number each time a scale was selected. In three cases where the answer was missing, we defaulted the answer to "I don't know".

In 86% of all cases, the WASHtsApp's answers were assessed as "perfect" or "sufficient". In 3% of the cases, answers were assessed as "wrong". However, interestingly, 10 out of 11 "wrong" assessments came from one expert, while the other experts assessed the same question-answer pairs as perfect or sufficient. Ambiguity in context-free WhatsApp messages may lead to varying assessments of the chatbot's responses.

---

[2] The "Healthy Village Facilitator's Guide," sourced online from the Solomon Islands context, was deemed highly relevant to our rural Ugandan setting by our experts.

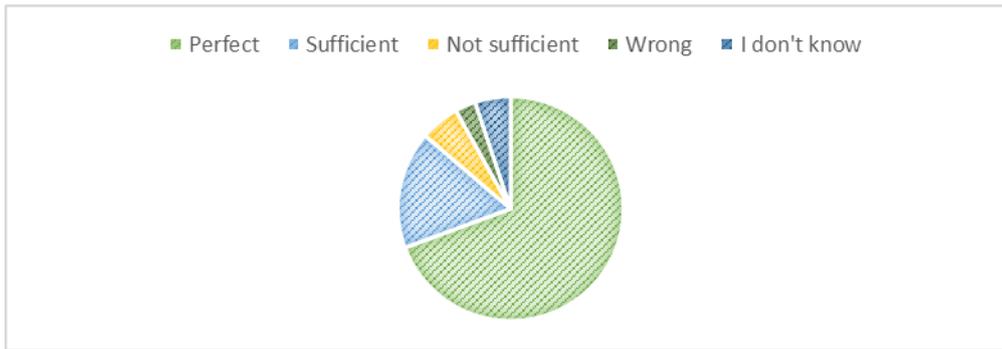

*Figure 2 Proportion of answer assessments by four experts.*

Additionally, experts were asked to assess the overall usability and suitability of the solution, which received positive feedback.

### 5.2 Community Validation

Seventy-seven persons completed the community validation questionnaire after interacting with WASHtsApp. Six respondents were removed from the evaluation, as their answers seemed unrealistically inconsistent within the items. Assuming a moderate effect size, a power level of 0.8 and a probability level of 0.05, a statistical power analysis would require 57 responses to detect the effect, indicating that the remaining 71 allow a statistically sound evaluation. 55% of respondents were aged 19-29, followed by the 30-39 group (28%). 62% were male. The results are presented in Table 3. Cronbach's alpha for all constructs exceeded the typical threshold of 0.7, and the correlations between the constructs align with previous TAM research and the theoretical assumptions.

*Table 3 Evaluation of Technology Acceptance Model questions of the community validation questionnaire. "Fully agree" was mapped to 5, "fully disagree" to 1. (\*\*\* indicates significance at the 0.001 level)*

| *Construct* | *Average* | *Standard Deviation* | *Cronbach's Alpha* | *Correlation with "Intention to Use"* |
|---|---|---|---|---|
| Perceived Usefulness | 4.11 | 1.06 | 0.74 | 0.45\*\*\* |
| Ease of Use | 4.36 | 0.86 | 0.77 | 0.58\*\*\* |
| Intention to Use | 4.34 | 0.92 | 0.87 | - |

The results clearly indicate a positive reception from potential users. This is also reflected in comments such as "My area is in dire need of safe water and sanitation facilities", which was the most common theme, mentioned eight times in the "Further comments" section at the end of the questionnaire. Six comments expressed gratitude to the app and developers, while four showed interest in future developments. At the same time, some users seemed to confuse WASHtsApp with WhatsApp, as four comments requested that the researchers develop a feature to download WhatsApp statuses. Four comments expressed concern about the chatbot's high mobile data usage. Two comments suggested supporting local languages, while another mentioned the challenge of reaching illiterate users.

Major limitations of both evaluations are the limited knowledge base the chatbot is currently working on and the fact that some features to support the most vulnerable groups, like local language support, etc., have not yet been implemented.

Overall, the evaluation supports the assumptions that WASHtsApp is both needed, helpful, and likely to be well-received by users.

## 6. Business Benefits

WASHtsApp does not directly serve a business use case and does not intend to sell its service for money. However, progress in SDG6 undoubtedly significantly contributes to a country's development and economic growth [1]. Reduced healthcare costs, decreased sick leave and income loss, and increased capacity, particularly among vulnerable populations, can contribute to both national GDP and individual economic growth.

Benefits for the users and communities will be driven mainly by changed behaviour due to increased knowledge about the causes and risks of the consumption of unsafe water, as well as strategies to improve water quality. In addition, users benefit from access to information becoming available independently of time and place.

Benefits for other stakeholders, primarily parties involved in providing safe water and respective education, benefit by accessing the data collected with the service to understand the needs of the communities further and can get in contact directly without the need for physical presence.

## 7. Conclusions

This paper discussed and evaluated WASHtsApp's applicability for educational and coaching deployments supporting SDG6, as suggested by, i.e., [10], [11]. We found that the chatbot's content quality and community acceptance were validated.
The following lessons learned can be formulated: We assume the approach is not to launch an app on our own but to use WhatsApp as a channel instead to succeed, as we gained some users who would not have installed a specific app. However, some users had trouble differentiating our service from WhatsApp as the app provider. Strategies should be placed for clear communication and more distinguishing in design. Secondly, we found that many respondents also expected immediate help from us. For many users, the service seemed to be an entrance door for more support. Future implementation must consider this to either have this help in place or communicate very clearly what further help users can expect to not disappoint users.

Indeed, the current prototype still leaves plenty of space to be further developed. The knowledge base materials could be tailored more closely to the local context. The knowledge base can be enriched with documents on local legislation, organisations, and initiatives promoting improved water access and sanitation. Additionally, we plan to support voice and image inputs and multiple local languages.

However, the potential usefulness of such a system is already apparent at that level and can be extended far beyond the described use case. Each user interaction with the chatbot deepens our understanding of current water, sanitation, and hygiene challenges the rural population faces. By leveraging WhatsApp's bidirectional communication and collected contact data, we can gather additional information on user demographics, such as gender, location, and age. This will further help to make informed decisions on which areas require intervention. Interventions can be implemented either through on-the-ground teams or by delivering tailored information via WASHtsApp.

Another potential application is rapid disease outbreak detection through field feedback. One exemplary scenario would be to ask all users on a regular basis "how many persons with typhus they currently know within their neighbourhoods". An accumulation of responses in a particular area would indicate that there is a potential source and intervention is needed. Once again, interventions could be implemented either through on-the-ground teams or by providing support to affected individuals via WhatsApp.

Subsequent Design Cycles will focus on these issues. The overall project's goal is to combine "Community Data Stewardship" with a positive form of the idea (kernel theory) of "Surveillance Capitalism" [21] to a non-profit use case that benefits society overall. Design

Science Research Cycle Two will enhance answer quality and expand the chatbot's reach through various online and offline channels to target the intended user group. This will be achieved in collaboration with local and international organisations promoting WASH principles in Uganda. Design Science Research Cycle Three will focus on developing a platform to leverage WASHtsApp's reach for additional purposes. The ultimate goal is to utilise the user base for educational purposes and to provide direct, timely information to affected communities.

In conclusion, the positive reception of WASHtsApp indicates its potential to address critical challenges related to SDG6. WhatsApp was a well-received channel, and RAG was deemed suitable for delivering localised and tailored WASH advice.

## Acknowledgements

We express our gratitude to our WASH experts, Rhodah Nassanga, Sulman Muhunguzi, Joshua Kurtz, and Lukas Dittrich, for their valuable contribution to the content validity assessment of this study. Their support, provided without financial compensation, is sincerely appreciated.

***Declaration of use of content generated by Artificial Intelligence (AI) (including but not limited to Generative-AI) in the paper***

The authors acknowledge the use of content generated by Artificial Intelligence (AI) (including but not limited to text, figures, images, and code) in the paper entitled "Enter Paper Title" in the following ways:
- Checking for grammar and typos.
- Suggestion of more concise formulations

# Online Appendix

https://drive.google.com/drive/folders/1oI05sNQUBYKPN_57_ZEIlH7EoAdY18iE?usp=sharing